\documentclass{article}

\usepackage[
  paperheight=8.5in,
  paperwidth=5.5in,
  left=10mm,
  right=10mm,
  top=20mm,
  bottom=20mm]{geometry}
\usepackage[utf8]{inputenc}

\usepackage{graphicx}
\usepackage{wrapfig}
\usepackage[bottom]{footmisc}
\usepackage{listings}
\usepackage{enumitem}

\usepackage{wrapfig}
\usepackage{ragged2e}

\usepackage{array}
\usepackage[table]{xcolor}
\usepackage{multirow}
\usepackage{booktabs}
\usepackage{hhline}
\definecolor{palegreen}{rgb}{0.6,0.98,0.6}

\usepackage{amsmath}
\usepackage{amssymb}
\usepackage{multicol}
\usepackage{lipsum}
\usepackage{hyphenat}
\PassOptionsToPackage{hyphens}{url}
\usepackage{url}

\usepackage{rotating}

\usepackage[T1]{fontenc}
\usepackage{textcomp}
\usepackage{listings}
\usepackage{setspace}

\newcommand*{\affmark}[1][*]{\textsuperscript{#1}}
\newcommand*{\email}[1]{\small{\texttt{#1}}}

\renewcommand{\footnoterule}{%
  \kern -3pt
  \hrule width \textwidth height 0.5pt
  \kern 2pt
}

\date{}

\usepackage{titlesec}
\titleformat*{\section}{\large\bfseries}
\titleformat*{\subsection}{\normalsize\bfseries}
\titleformat*{\subsubsection}{\normalsize\bfseries}

\usepackage{biblatex}

\title{Carry-Over Lottery Allocation: Practical Incentive-Compatible Drafts}

\author{
Timothy Highley\affmark[1], Tannah Duncan\affmark[1], and Ilia Volkov\affmark[1]\\
\affmark[1]Department of Mathematics and Computer Science\\
La Salle University\\
Philadelphia, PA 19141\\
1-215-951-1722\\
\email{\{highley,duncant4,volkovi1\}@lasalle.edu}\\
}

\begin{document}
\maketitle

\begin{abstract}
The NBA draft can incentivize teams to deliberately lose. We propose a draft mechanism that is practical, incentive-compatible, and favors weaker teams. The Carry-Over Lottery Allocation (COLA) framework represents a paradigm shift in evaluating team quality, replacing single season standings with multi-year playoff outcomes. In our proposed mechanism, every non-playoff team receives the same number of lottery tickets, removing incentives to lose. Lottery tickets carry over to future lotteries, but playoff success or winning a top pick diminishes a team’s accumulated tickets. The lottery is familiar and preserves fan engagement.  

Implementation challenges are addressed to demonstrate feasibility, including transitioning to COLA, handling trades, and accommodating draft classes of varying strength. For exceptionally strong classes, teams may prefer the lottery to the playoffs. We provide a solution, employing a truth-elicitation mechanism to identify such years and expanding  lottery eligibility to include as many playoff teams as necessary to preserve incentive compatibility.
\end{abstract}

\section{Introduction}

Every NBA season, some of the league's worst teams face a perverse incentive: intentionally losing games (tanking) can improve their draft prospects. Tanking undermines competitive integrity, frustrates fans, and contradicts the fundamental premise of professional sports. 

In the current NBA draft system, all non-playoff teams enter a lottery for the top four draft positions, with worse records giving higher chances of winning. Teams that do not win a top pick are then slotted by record, starting with the worst team. This structure creates a trade-off: the league wants to reward weaker teams to maintain competitive balance, but higher lottery odds for the worst teams also increase incentives to lose. Despite repeated adjustments to lottery probabilities over the decades, tanking persists because of a fundamental impossibility: Munro and Banchio \cite{munro_banchio} proved that no draft mechanism based solely on end-of-season records can simultaneously reward poor performance and eliminate incentives to tank. The current system is, mathematically, doomed to fail.

From a mechanism-design perspective, the challenge is clear: losing games should never maximize a team's expected utility. While numerous alternatives have been proposed, each one has trade-offs, and no practical proposal simultaneously rewards the weakest teams while fully eliminating incentives to tank. This leaves a central question: does a practical, fully incentive-compatible mechanism exist, and if so, what design choices are necessary to achieve it?

We answer in the affirmative with the proposal of the \textbf{Carry-Over Lottery Allocation (COLA)} framework. COLA achieves incentive compatibility by eliminating any advantage from additional losses. Each team maintains a stockpile of tickets that grows and shrinks, but the stockpile never grows more because of additional losses. COLA rewards the weakest teams through carry-over: lottery tickets that do not win a top draft pick are retained for future lotteries, while playoff success or winning a top pick diminishes a team's accumulated tickets.

COLA is a framework, and there are many possible variants within the framework. This paper discusses possible variants later, but primarily focuses on Classic COLA, which is one instance within the framework.

Classic COLA is practical to implement. It uses a familiar draft with lottery, requiring minimal changes to existing league procedures and fan expectations. Traded draft picks need special handling, but COLA solves this by excluding traded picks from lottery eligibility. The mechanism also accommodates variations in draft class strength. For a weak class, a team might view a lottery win as undesirable due to a loss of tickets in future lotteries, so there is an option to skip the lottery for a cost. For a strong draft class, many teams might prefer to be in the lottery instead of the playoffs. This violates a key assumption of anti-tanking mechanisms: reaching the playoffs is the primary goal, with draft picks as a secondary consideration. This problem has no solution in existing mechanisms, because if all teams prefer the lottery, no mechanism can both reward weak teams and prevent tanking. To address this, COLA identifies strong draft classes in advance using a Bayesian Truth Serum–based media survey and correspondingly expands lottery access as far as necessary but only when necessary, maintaining incentive compatibility without altering the core structure of the mechanism.

The remainder of this paper proceeds as follows. In Section \ref{sec:RelatedWork}, we review related work, including other draft mechanisms. Section \ref{sec:MainProposal} presents the core mechanism: how lottery tickets accumulate, how they are diminished by playoff success or lottery wins, and why this structure ensures that teams never maximize their expected utility by losing. Section \ref{sec:Implementation} examines critical implementation challenges, including the handling of traded draft picks, transitioning from the current system, and addressing potential violations of our preference assumptions. Section \ref{sec:Simulation} presents simulation results that provide further evidence that under COLA all teams have an equal likelihood of success over time. In an appendix, we provide a formal proof that there exists a draft mechanism that is incentive-compatible and favors the weakest team.

\section{Related Work} \label{sec:RelatedWork}

\subsection{Draft Mechanism Review}

There are three important criteria for a draft mechanism:

\begin{enumerate}
    \item \textbf{Practicality}. Can the mechanism be implemented in the real world without creating political, fan-facing, or operational issues? Do its assumptions align with real-world conditions? Mechanisms that are too complex, radical, rely on unrealistic assumptions, or require precise knowledge of team preferences fail this criterion.
    \item \textbf{Anti-Tanking.} Deliberately losing games should not improve a team's expected utility under reasonable assumptions. Mechanisms may still be theoretically Anti-Tanking, but if the underlying assumptions fail in practice, the mechanism may become impractical.
    \item \textbf{Preferencing for Quality} (or simply \textbf{Preferencing} in this section). For the sake of competitive balance, the mechanism should favor the worst teams by increasing their chances of obtaining high draft picks relative to better teams.
\end{enumerate}

We now review several proposed mechanisms according to those criteria:

\begin{itemize}
    \item \textbf{Munro and Banchio} \cite{munro_banchio} propose that the lottery be based on standings at a dynamically-determined point in the season. It meets the Preferencing and Anti-Tanking criteria under standard assumptions, but does not meet the Practicality criterion for two reasons. First, the calculation of the dynamically-determined date is opaque to a casual fan and requires knowledge of private team preferences. Choosing a predetermined early cutoff date may seem to approximate their approach, but if the date is announced and does not take into account teams' incentives, then it simply moves tanking to earlier in the year. Second, in most seasons there are teams that believe their likelihood of playoff success is so small that they have an incentive to tank as soon as the season starts, in which case there are no standings on which to base the lottery.
    
    \item \textbf{The WNBA} bases the lottery on the combined record of the two most recent seasons. This improves Preferencing for persistently weak teams but it is not an Anti-Tanking mechanism. A loss in one season could affect the draft position for two seasons, giving even playoff teams incentives to underperform.
    
    \item \textbf{Gold Plan} \cite{goldplan2012} bases the draft on the number of wins that a team earns after playoff elimination. Under the assumption that every team puts forth maximum effort until they are mathematically eliminated, it satisfies Anti-Tanking. However, in practice it is common for teams to tank before they are eliminated. This mechanism can incentivize tanking earlier in the season so that they are eliminated as soon as possible. That way, the team has more time to rack up post-elimination wins, so due to practical concerns The Gold Plan is not truly Anti-Tanking. In terms of Preferencing, some weaker teams are favored over some stronger teams, but the Gold Plan disadvantages truly bad teams that cannot win many games even if they are eliminated early.
    
    \item \textbf{Uniform randomization} gives every team an equal likelihood of every draft position. \textbf{The Wheel} \cite{lowe2013nba_wheel} is a proposal where there is a thirty-year rotation of teams taking turns in each draft position without regard for standings or playoff success. Both of these proposals give no advantage at all to any team, so they are clearly Anti-Tanking approaches, but they completely eschew Preferencing.
        
    \item \textbf{Auctions for draft picks} have been proposed in various formats, but they face  pushback from league decision-makers, making them impractical. There are many possible auction mechanisms, some of which would satisfy Preferencing and Anti-Tanking. For example, an auction could be paired with COLA to disperse auction currency instead of lottery tickets. That would meet the Preferencing and Anti-Tanking criteria. Practicality rules all of them out in an NBA context.
    
    \item \textbf{Relegation-style systems} such as those found in European soccer (football) leagues are politically infeasible with NBA ownership, thus failing Practicality.
    
    \item \textbf{Explicit punishment of tanking} fails Practicality, because it is not possible to reliably prove intent. It also risks punishing already weak teams and uneven enforcement, which would produce bad optics.
\end{itemize}

In summary, none of the existing approaches satisfy all three criteria simultaneously. The COLA Draft Mechanism is designed to meet all three objectives, providing a practical, anti-tanking system that appropriately advantages teams with the poorest performance.

\subsection{Truth Elicitation}

A key challenge in implementing an incentive-compatible draft mechanism is determining when teams might prefer lottery access to playoff participation. When exceptionally strong prospects enter the draft, even successful teams may prefer missing the playoffs. We employ a truth-elicitation mechanism that aggregates expert assessments of draft class strength and the resulting competitive incentives, ensuring honest reporting even when there is no objective way to verify the answers.

The Bayesian Truth Serum (BTS) \cite{prelec2004bayesian} was presented in a groundbreaking paper that showed that, when requesting binary signals from large populations, agents can be incentivized to report their true beliefs by soliciting both their own beliefs and a prediction about others' reports. Robust BTS \cite{witkowski2012robust} extends this to small populations but still assumes binary signals. Non-Binary RBTS \cite{radanovic2013nonbinary} further generalizes the method to handle multi-category signals, making it well-suited to our scenario, where we anticipate fewer than 100 media respondents and a six-point scale to measure the severity of the tanking problem in a given season. 

Other peer-prediction approaches highlight complementary ideas: Miller et al. \cite{miller2005eliciting} demonstrate eliciting truthful probabilistic forecasts with proper scoring rules; Roth and Schoenebeck \cite{roth2012conducting} discuss how to minimize the cost of incentivizing a representative sample of a population to participate in a survey; and Radanovic, Faltings, and Jurca \cite{radanovic2016incentives} explore a model where responses to multiple tasks or questions are used to verify that an agent is exerting effort. Lehmann \cite{lehmann2024mechanisms} provides a more comprehensive overview of the landscape of truth elicitation mechanisms.

Some of these methods require large populations, multiple questions/tasks, or binary signals. Our scenario does not meet those requirements, but it satisfies all of the requirements of the RBTS variant we adopt.

By incorporating RBTS, we can reliably extract media assessments of what tanking incentives teams face, and use that information to calibrate the incentive-compatible structures.

\section{An Incentive-Compatible Draft Mechanism} \label{sec:MainProposal}
\subsection{A New Paradigm for Sports Drafts} \label{sec:Paradigm}

The impossibility result of Munro and Banchio \cite{munro_banchio} motivates us to reconsider how to determine which teams deserve the benefit of higher draft picks. This leads to a paradigm shift that has two parts.

Traditionally, teams with the worst records receive the highest draft picks (or the best odds of winning the highest draft picks). This makes sense, because it provides better competitive balance in a league if the best incoming players go to the worst teams. It helps to keep fans of all teams interested in the league. However, the goal should not be to give higher draft picks to the teams with the worst records. Instead, the goal should be to give those picks to the teams that actually are the worst. Teams with bad records tend to be bad teams, but the worst team and the team with the worst record are not necessarily equivalent. 

The first part of the paradigm shift is to change how team quality is measured. There is no better objective measure of team quality than team performance, but the regular season record (especially for just a single season) is a poor metric to use. Its use inherently incentivizes tanking. Multi-year performance and playoff success are more robust measures of team quality. This is especially true in the current context, where playoff success is valued so highly that many fans and analysts view the regular season as almost worthless by comparison. The important questions to ask, then, are:
\begin{itemize}
    \item Did the team make the playoffs?
    \item How many playoff series did the team win?
    \item What has been the team's playoff success in recent years?
\end{itemize}
Regular-season record does not enter the equation at all, except for the fact that regular-season record determines whether a team makes the playoffs.

The second part of the paradigm shift is to favor unfortunate teams when it comes to lottery picks. There are two ways in which teams may be considered fortunate: playoff success and winning high lottery picks. If all else is equal, teams that have had neither playoff success nor high lottery picks in recent years should be favored above teams that have had one or the other (or both).

We previously defined Practicality, Anti-Tanking, and Preferencing for Quality as important criteria for a draft mechanism. We now introduce two additional, important considerations:
\begin{itemize}
    \item \textbf{Preferencing for Luck.} In the interest of fairness, among teams of similar weakness, those that have had the worst luck in previous lotteries should be rewarded with higher draft picks than those that have been luckier.
    \item \textbf{Playoff Track Record.} Long-term playoff success is a better measure of team quality than single-year regular season records.
\end{itemize}

None of the proposals discussed earlier incorporate Preferencing for Luck or Playoff Track Record. (There have been recent discussions about prohibiting lottery winners from winning again the following year. That tweak would be a form of Preferencing for Luck.)

With this paradigm shift, we can design a draft mechanism that meets all five of the desired criteria. The following section presents Classic COLA, a practical mechanism that implements these principles.

\subsection{COLA Overview}

We introduce the Carry-Over Lottery Allocation (COLA) framework, named for two of its key features: draft picks are allocated via lottery, and some lottery tickets carry over between years. While there are many possible variants in this framework, we will focus on one instance that we will call Classic COLA. For brevity, we refer to this simply as COLA throughout the paper, reserving the ``Classic'' distinction for contexts where it is necessary to differentiate it from other variants. We note that COLA thus refers both to the overall framework and to this specific instance, with the meaning clear from context.

Each year, some NBA teams qualify for the playoffs, while others do not. Under COLA, the teams that fail to make the playoffs are entered into a lottery for the top four draft picks, which lines up with the current NBA draft system. We refer to each chance in the lottery as a lottery ticket. Each team has a lottery index, which is a number indicating how many lottery tickets they will have in the raffle for the top draft picks. 

COLA is based on three core ideas. First, to be Anti-Tanking, each year every non-playoff team’s lottery index is incremented by the same amount so that additional losses yield no advantage. Second, lottery indices carry over from year to year. Third, playoff success and lottery wins diminish a team’s index, achieving Preferencing for Quality and Luck.

We will use $\alpha$ to represent the amount added to the lottery index of each non-playoff team. COLA works for any value of $\alpha$, as long as it is large enough that rounding does not become a significant issue when a team's lottery index is diminished. We will use $\alpha = 1000$ as our assumption throughout the paper, and we will also assume lottery indices are integers. The number 1000 is large enough to capture the important differences between teams as values are diminished, but small enough that fans will be able to talk about the number of tickets each team has in the lottery without having numbers so large that they are a hindrance to conversation.

After the lottery, the lottery index of a team that did not win a top draft pick remains unchanged, which means that if they are in the next year's lottery they will receive more lottery tickets. The lottery index of a team that won a top draft pick will be diminished by some amount, based on which pick they received. Incentive compatibility is maintained with or without these rules for diminishing a team's lottery index. The numbers here are chosen in the interest of fairness and simplicity. In terms of fairness, whoever wins the top pick is more fortunate, so based on the Preferencing for Luck principle, the team that wins the first pick should lose more than the teams that win the other top draft picks. In terms of simplicity, we chose simple values for how much the totals are diminished. That way, they are easy to calculate, understand, and discuss:

\begin{itemize}
    \item The lottery index of the team that wins the first draft pick is reduced to 0. 
    \item The lottery index of the team that wins the second draft pick is diminished by 75\%. 
    \item The lottery index of the team that wins the third draft pick is diminished by 50\%.
    \item The lottery index of the team that wins the fourth draft pick is diminished by 25\%.
\end{itemize}    

Teams that make the playoffs also have a lottery index, but they do not receive lottery tickets or participate in the lottery. The lottery index is diminished based on playoff success:

\begin{itemize}
    \item The lottery index of the champion is reduced to 0. 
    \item The lottery index of the runner-up is diminished by 75\%. 
    \item The lottery index of a team that loses in the conference finals is diminished by 50\%. 
    \item The lottery index of a team that loses in the second round is diminished by 25\%. 
    \item The lottery index of a team that loses in the first round remains unchanged, but they do not participate in that year's lottery.
\end{itemize}

The diminishment of lottery indices for successful teams is not necessary to maintain incentive-compatibility. It is included in the interest of the Preferencing for Quality and Playoff Track Record principles. As with the diminishment rates for winning top draft picks, the numbers were chosen for simplicity.

\subsection{Incentive Compatibility}

We first establish COLA's incentive compatibility. In the next section, we discuss the reasonableness of the underlying assumptions and how to handle situations where they might fail. COLA is incentive-compatible if the following assumptions hold:
\begin{enumerate}
    \item In the regular season, every team has a primary goal of making the playoffs, a secondary goal of improving through the draft lottery, and a tertiary goal of winning games simply for pride and/or marketing.
    \item In the playoffs, every team's primary goal is advancing in the playoffs.
    \item The difference between the 5th through 14th draft picks is small enough to be considered negligible.
    \item The benefit from strategically influencing which teams participate in the lottery pool is sufficiently small that teams do not pursue it.
    \item Teams have not traded away draft picks with protections.    
\end{enumerate}

Consider teams playing three different types of games: regular season games while in playoff contention, regular season games after elimination, and playoff games.

Before elimination from playoff contention, winning increases the likelihood of accomplishing the primary goal (making the playoffs). Thus, a team has a positive incentive to win each game as long as it is still possible to make the playoffs. Teams that miss the playoffs automatically accomplish the secondary goal (lottery participation), so there is no incentive to pursue it, even if the probability of accomplishing the primary goal is vanishingly small. 

After elimination, a team cannot increase its own lottery index beyond what is automatically granted for missing the playoffs. All non-playoff teams are treated equally, and by Assumption 3 the difference between the 5th and 14th picks is negligible, so losing additional games yields no meaningful draft advantage.

A team's probability of winning the lottery is
\[
\text{Chance of winning} = \frac{\text{team's lottery tickets}}{\text{total lottery tickets}}.
\]

After elimination, the numerator is fixed, and the only way to improve the fraction is by influencing which other teams enter the lottery (the denominator). By Assumption 4, any potential gain from this is too small to pursue. Therefore, eliminated teams will win games for pride rather than strategically lose.

In the playoffs, the lottery index of a team that advances will be diminished, but as long as winning in the playoffs is the primary goal (i.e. the second assumption holds), teams will try to win.

The above analysis assumes that teams have not traded away their draft picks with protections on them. Under the stated assumptions, the Anti-Tanking property is met. Each of the assumptions is necessary, so we now consider whether each is reasonable.

\subsection{Underlying Assumptions} \label{sec:Assumptions}

The first assumption is usually true. Most teams would prefer to make the playoffs instead of missing the playoffs. The second assumption has to do with tanking during the playoffs, which is something we have never seen. Under COLA, teams that advance in the playoffs lose more lottery tickets, which could potentially create an incentive to lose games even in the playoffs. However, because playoff success is the primary goal of a team, it seems unlikely. If these assumptions were to be violated, it would be when there is a strong draft class. Therefore, it is important to identify when such a situation might arise. Both of the first two assumptions can be addressed through the application of a Bayesian Truth Serum, which we will cover in Section \ref{sec:StrongDraft}. 

The third assumption has been true historically. For example, we do not see teams tanking just for the expectation of moving from a 13th pick up to the 12th pick. If it is determined that the difference between the 5th through 14th picks is significant enough to induce tanking, then the solution is to also raffle off those picks through the lottery.

The fourth assumption addresses an edge case where a lottery team could marginally improve its chances of winning the lottery by losing to a high-index opponent near the playoff line, swapping them out of the lottery pool, replacing them with a low-index opponent. The situation requires specific conditions: the team must be eliminated from playoff contention, two other teams near the line must have substantially different lottery indices, and the game outcome must actually shift the standings. Simulations indicate that teams can expect their benefit from this scenario to be less than 1\% (see Section \ref{sec:Simulation}). This is akin to tanking to move from the 10th pick to the 9th, which is behavior we do not observe in practice. Moreover, the play-in tournament and uncertainty in other game outcomes make successfully executing such a strategy highly unreliable. If this were a practical concern, then including first-round playoff losers in the lottery would eliminate the potential for non-playoff teams to determine whether other teams are in the lottery.

The fifth assumption is simply false, unfortunately. Teams do trade away picks with protections, and that introduces much stronger incentives to tank. While the ability to go from an 8th pick to a 7th pick may be an insignificant reason to tank, the possibility of going from no pick at all to a top draft pick is a strong incentive, and there are apparent examples of exactly that in recent years. We will discuss this issue in more detail later, but one simple solution is to simply disallow protections on traded draft picks.

The incentive analysis above establishes the Anti-Tanking property of COLA. The worst teams accrue more tickets than the best teams, satisfying Preferencing for Quality by using Playoff Track Record as a measurement of team quality. Teams unlucky in the lottery gain future advantage, satisfying Preferencing for Luck. Practicality is addressed through implementation details discussed below.

In the appendix, we provide a formal proof that, under reasonable assumptions, there exists a variant of COLA that is incentive-compatible and gives the weakest team the best odds of winning the first draft pick. Classic COLA satisfies the first property but not the second. Because it favors both weak and unlucky teams, the weakest team is not guaranteed to have the best odds of winning the first pick.

\section{Implementation Details} \label{sec:Implementation}

\subsection{Transition}
The transition from the current NBA draft to the COLA system is straightforward: determine a starting lottery index for each team. Every NBA team has either had a top two NBA draft pick or made the NBA finals at least once since 1999, resulting in a substantial lottery index diminishment for every franchise since then.

Therefore, with 1998-99 as our starting year, we are able to capture all recent, meaningful history. We apply lottery index increments and diminishments each year as appropriate to determine each team's current lottery index. (We apply the 25\% decrement to \#4 picks each year, even before they were included as part of the lottery.) Table \ref{tab:lottery_index_each_team_simple} shows what each team's lottery index was at the end of the 2025 regular season, along with the probability they would have won the number one pick under COLA. For comparison, their actual probability under the current system is also listed. The final column shows the lottery indices after post-draft and post-playoff adjustments.

\begin{table}[ht]
\centering
\caption{Lottery Index for Each Team}
\label{tab:lottery_index_each_team_simple}
\resizebox{\textwidth}{!}{
\begin{tabular}{|l|r|r|r|r|r|}
\hline
\textbf{Team} &
\shortstack{\textbf{Lottery}\\\textbf{Index}\\\textbf{(1 May 25)}} &
\shortstack{\textbf{Lottery}\\\textbf{Tickets}} &
\shortstack{\textbf{Win Prob.}\\\textbf{(COLA)}} &
\shortstack{\textbf{Win Prob.}\\\textbf{(Historical}\\\textbf{Draft)}} &
\shortstack{\textbf{Lottery}\\\textbf{Index}\\\textbf{(1 Sep 25)}} \\
\hline
Sacramento Kings & 7109 & 7109 & 14.4 & 0.8 & 7109 \\
Chicago Bulls & 7000 & 7000 & 14.2 & 1.8 & 7000 \\
Brooklyn Nets & 6226 & 6226 & 12.6 & 9.0 & 6226 \\
Utah Jazz & 5018 & 5018 & 10.1 & 14.0 & 5018 \\
New Orleans Pelicans & 4000 & 4000 & 8.1 & 12.5 & 4000 \\
New York Knicks & 3938 & 0 & 0 & 0 & 1969 \\
Toronto Raptors & 3750 & 3750 & 7.6 & 7.5 & 3750 \\
Charlotte Hornets & 3612 & 3612 & 7.3 & 14.0 & 2709 \\
Portland Trail Blazers & 3422 & 3422 & 6.9 & 3.8 & 3422 \\
Detroit Pistons & 3000 & 0 & 0 & 0 & 3000 \\
Washington Wizards & 2543 & 2543 & 5.1 & 14.0 & 2543 \\
Indiana Pacers & 2375 & 0 & 0 & 0 & 594 \\
Memphis Grizzlies & 2166 & 0 & 0 & 0 & 2166 \\
Cleveland Cavaliers & 1875 & 0 & 0 & 0 & 1406 \\
Dallas Mavericks & 1750 & 1750 & 3.5 & 1.8 & 0 \\
San Antonio Spurs & 1750 & 1750 & 3.5 & 6.0 & 438 \\
Los Angeles Clippers & 1692 & 0 & 0 & 0 & 1692 \\
Phoenix Suns & 1281 & 1281 & 2.6 & 3.8 & 1281 \\
Oklahoma City Thunder & 1225 & 0 & 0 & 0 & 0 \\
Houston Rockets & 1132 & 0 & 0 & 0 & 1132 \\
Golden State Warriors & 1000 & 0 & 0 & 0 & 750 \\
Orlando Magic & 1000 & 0 & 0 & 0 & 1000 \\
Atlanta Hawks & 1000 & 1000 & 2.0 & 0.8 & 1000 \\
Philadelphia 76ers & 1000 & 1000 & 2.0 & 10.5 & 500 \\
Los Angeles Lakers & 500 & 0 & 0 & 0 & 500 \\
Minnesota Timberwolves & 500 & 0 & 0 & 0 & 250 \\
Miami Heat & 86 & 0 & 0 & 0 & 86 \\
Boston Celtics & 0 & 0 & 0 & 0 & 0 \\
Denver Nuggets & 0 & 0 & 0 & 0 & 0 \\
Milwaukee Bucks & 0 & 0 & 0 & 0 & 0 \\
\hline
\end{tabular}}
\end{table}

The team with the highest lottery index is the Sacramento Kings, due to their long run of bad luck in the draft along with little to no playoff success. They had no playoff series wins from 2005-2025, and during that time had just three lottery wins (a \#2 pick in 2018 and two \#4 picks along the way).

Relative to the 2025 historical lottery, the three teams with the greatest drop in win probability under COLA are the Wizards (-8.9\%), Sixers (-8.5\%), and Hornets (-6.7\%). It is easy to see why. The Wizards and Hornets won the \#2 pick in 2024 and 2023, respectively. The Sixers had made the playoffs each of the previous seven years, with five playoff series wins. (The winners of the 2023 and 2024 \#1 picks did not have a very high win probability under either COLA or the historical lottery.)

\subsection{Traded Picks}

Traded draft picks require careful handling under COLA. Banning them entirely would be a simple way to maintain incentive-compatibility, but it is unnecessary and also impractical, because it would reduce opportunities for team improvement and diminish fan interest.

A natural approach would be to diminish the team holding the pick when it wins the lottery, since they receive the benefit. However, we reject this option because it creates a perverse incentive: teams could repeatedly trade away their own valuable high-lottery-index picks to avoid diminishment while continuing to accumulate lottery tickets indefinitely. Diminishing the original team as well would close the loophole, but then the original team would have a lowered index while gaining nothing from a pick it no longer owns. This creates a risk that short-term decision-makers mortgage the future. We elect not to allow this for the same reason the Stepien rule exists.

Instead, we prohibit a team from winning the lottery using a draft pick that they traded for. When a draft pick is traded, the original team may trade it either without any protections or with protections for picks 1-4. No other protections are permitted. If the pick is traded without protections, it is excluded from the draft lottery entirely, and it will be slotted as usual in draft position 5+. If the pick is traded with protections, then if the draft pick is chosen as a lottery winner, it reverts back to the original team, and the original team's lottery index is diminished accordingly.

These rules on traded picks maintain incentive compatibility. A team that has traded away its pick without protections has no incentive to tank, because losing only helps the team that holds it. A team that does not own its own pick and has protections 1-4 has no incentive to lose after playoff elimination, because the team can do nothing additional to make a lottery win more likely. The team may have more motivation to miss the playoffs entirely if a draft pick has protections 1-4. However, under the assumption that the playoffs are the primary goal, they still have no incentive to lose. (Section \ref{sec:StrongDraft} addresses the situation where teams actually do want to miss the playoffs.) 

Protections beyond picks 1-4 would create tanking incentives, as the original team could lose strategically to ensure the pick falls within the protected range. Although draft picks that have already been traded with additional protections must be honored, the practice would be banned going forward to prevent tanking.

\subsection{Weak Draft Classes}
If a team wins the lottery during a year when the draft class is weak, they lose a portion of their lottery index but do not necessarily improve. Winning the lottery becomes an unfortunate outcome: a team with a large lottery index sees it drastically reduced, leaving them with worse chances in future years when draft classes are stronger. To address this, teams may opt out of the lottery entirely and accept their draft position as if they had not won a top pick. Doing so costs 2000 lottery index points, which is equal to $2\alpha$, or two years' worth of increments.

The penalty for skipping is steep because broad participation benefits the league. Lottery participation generates fan interest, is a primary mechanism for weak teams to improve, and avoids the insult to draft-eligible players when teams publicly opt-out of their class. If many weak teams skip the lottery, top picks go to stronger teams. That harms competitive balance. With a 2000-point penalty, only teams with very large lottery indices facing exceptionally weak draft classes would consider opting out. This is expected to be rare. A team facing this situation would be more likely to trade the pick away without protections than to exercise an opt-out, but the opt-out provides a failsafe and places a floor on the value of the draft pick.

COLA meets all of the desirable criteria with or without this opt-out provision. This provision exists only because carry-over could otherwise create situations where winning the lottery reduces a team's utility. Teams should always welcome a lottery win, so we provide teams this option to avoid that outcome, even if they are unable to find an amenable trade partner.

\subsection{Strong Draft Classes} \label{sec:StrongDraft}
COLA is incentive-compatible under the assumption that teams prefer making the playoffs over entering the draft lottery. In years with strong draft classes, some teams might prefer missing the playoffs in order to secure a chance at a top pick. If most teams still compete for playoff spots, those few teams are likely eliminated through normal competition, achieving their goal without intentional losses. Once eliminated, all non-playoff teams receive the same lottery index increment, eliminating any incentive to lose additional games. Thus, problematic tanking arises only if teams that would otherwise make the playoffs prefer the lottery. Strong draft classes represent the scenario where this foundational assumption is most likely to fail in ways that matter for incentive compatibility.

When the assumption does fail, removing tanking incentives requires adjusting the line between lottery and non-lottery teams. Normally, this line falls at playoff qualification: playoff teams are excluded while all non-playoff teams participate in the lottery and receive equal lottery index increments. In order to maintain incentive compatibility in years with exceptionally strong draft classes, the line must be repositioned to include some playoff teams in the lottery.

This creates a fundamental tension. For the purposes of Anti-Tanking, the line must be repositioned to eliminate bad incentives. However, for the purposes of Preferencing for Quality, the line should remain at playoff qualification, ensuring only non-playoff teams benefit from the lottery. These goals conflict: raising the line to include playoff teams removes tanking incentives but dilutes the advantage for truly weak teams.

We need a way to accurately assess whether teams will tank to miss the playoffs, but we cannot rely on teams themselves to provide this information. Teams have strong incentives to misrepresent their intentions. Beyond strategic considerations, teams value their reputation for integrity and competitive drive, and that value is impossible to quantify or incorporate into an incentive mechanism.

Asking teams to report on \textit{other} teams' likely behavior is equally problematic. Teams expecting to make the playoffs would claim that tanking is widespread and the line should be raised, allowing them to participate in the lottery. Teams expecting to miss the playoffs would claim that no tanking will occur and the line should remain unchanged, excluding playoff teams and improving their own lottery odds. Both groups would have an incentive to make these claims regardless of their true beliefs. Quantifying and correcting for those incentives would require detailed knowledge of how each team values potential draft picks, which is not feasible.

One alternative would be for the league office to decide unilaterally whether a draft class is strong enough to move the line. While simple, this approach is unattractive. It places a single, high-stakes decision in the hands of one person, with no objective standard and no way to determine afterward whether the decision was correct. If the line is moved and a playoff team receives the top pick, the legitimacy of the decision will be questioned. If the line is not moved and tanking does occur, the league will be blamed. Even if made in good faith, the decision would appear arbitrary and invite controversy.

The league office also faces considerable downside if something goes wrong with essentially no upside to getting it right. If the line is moved and doing so was necessary, there is no clear way to demonstrate that afterward. If the line is not moved and tanking would not have occurred anyway, the outcome is indistinguishable from the status quo. In contrast, the downside of getting the decision wrong is substantial. Moving the line unnecessarily risks awarding a top draft pick to a strong team, while failing to move the line when needed allows tanking to undermine competitive integrity. Putting this decision on one person is therefore undesirable.

A collective assessment is more likely to be accurate, because it aggregates information from many independent observers about draft class strength and teams’ incentives. Just as importantly, it is more likely to be accepted. Distributing the decision across a group diffuses responsibility, reduces reputational and political risk for the league, and avoids the appearance of arbitrary intervention. Therefore, we rely on external experts (e.g., media and analysts) to assess teams’ incentives to tank, using an incentive-compatible truth-elicitation mechanism to ensure honest reporting.

\subsubsection{Bayesian Truth Serum}

The Robust Bayesian Truth Serum (RBTS) by Radanovic and Faltings \cite{radanovic2013nonbinary} can be used to accurately assess whether teams will tank to miss the playoffs, and also whether they might try to lose even during the playoffs. We can use it to determine where the line needs to be in a particular year so that there is no incentive for teams to lose. RBTS is a truth-elicitation mechanism that incentivizes honest reporting of private beliefs even when ground truth cannot be directly observed. Rather than attempting to calculate teams' incentives, which would require quantifying immeasurable preferences like how a team values a championship versus a potentially franchise-altering draft pick, we aggregate expert assessments through media surveys. RBTS provides the incentive structure that ensures these experts report their true beliefs, allowing us to reliably determine where to set the line between lottery participants and non-participants.

\subsubsection{Suitability of Bayesian Truth Serum for the NBA Draft}
The parameters of this scenario fit the requirements of RBTS. Our scenario requests a non-binary signal from a relatively small population, and it is requesting only one signal from each agent. Those restrictions eliminate many other truth elicitation algorithms as options \cite{miller2005eliciting} \cite{prelec2004bayesian} \cite{witkowski2012robust} \cite{zhang2025stochastically}, but RBTS works under those parameters.

RBTS requires that agents be able to expect other agents to answer honestly. To meet this requirement, votes will be collected via secret ballot to prevent collusion, but after results are tallied, each media member's vote and written explanation will be published. This accountability structure prevents obviously dishonest reporting: media members cannot make brazenly false claims without damaging their credibility. However, publication alone does not prevent conformity. A media member with an unpopular opinion or one who does not wish to take the time to do a proper analysis might report what they expect others to say rather than their true assessment. RBTS addresses this by rewarding accurate, honest reporting even when an agent's belief differs from the majority. While RBTS ensures that honesty is an equilibrium, other equilibria may exist. The public accountability structure ensures that the honest equilibrium is selected. Media members who provide misleading assessments will be excluded from future surveys. Exclusion is based on demonstrably bad-faith behavior rather than disagreement with outcomes. Together, these mechanisms incentivize participants to report their true beliefs.

RBTS requires that the self-predicting assumption be met. Each media member reports a signal, which is their assessment of where the line should be drawn to prevent tanking. The self-predicting assumption can be stated as, ``An agent expects their own signal to be most common among other agents when they themselves hold that belief.'' For example, if a media member believes the line should be moved to include second-round playoff losers, they might predict that 20\% of other media members will report the same. However, suppose that media member held a different belief instead. Perhaps they they think that no line movement is necessary. In that case, they would expect a lower percentage (perhaps 5\%) would call for moving the line to the second round. Put differently: for any given signal, an agent expects it will be most prevalent among others when they themselves observe that signal.

Note that this does not require the agent to believe their signal is the most common response overall. A media member might expect only 20\% to say ``second round'' while 60\% say ``don't move the line.'' The assumption only requires that the 20\% is higher than what they would expect if they held a different belief. This assumption holds naturally in our context because media members analyze similar information about draft class strength, team incentives, and playoff probabilities. When a media member reaches a particular conclusion based on this shared information, they expect other informed analysts to reach similar conclusions more often than they would if the media member had reached a different conclusion instead.

\subsubsection{The Annual Survey}

With the suitability of RBTS established, we now detail how the survey works in practice. Each year before the season, selected media members and analysts participate in a survey. The survey is a single multiple-choice question with six options: Does the line need to be moved in order to prevent tanking in the NBA this year?

\begin{enumerate}
    \item No team will tank. There is no need to move the line.
    \item Move the line to include teams that lose in the first round of the playoffs.
    \item Move the line to include all teams that lose in the first two rounds of the playoffs.
    \item Move the line to include all teams that lose in the conference finals or earlier.
    \item Move the line to include all teams that lose in the NBA Finals or earlier. That is, all teams except the NBA champion.
    \item Move the line to include all teams, including the NBA champion, as part of the draft lottery.
\end{enumerate}

Additionally, each agent predicts the outcome of the poll: what percentage of the responses will each of the options receive? Their survey response and prediction are required for the RBTS mechanism.

Set the line based on the results of this poll as follows: move the line to the furthest point where at least 50\% of respondents indicated the line should be moved to that position or beyond. For example, if 30\% say ``first round,'' 40\% say ``second round,'' and 30\% say ``don't move,'' then 70\% support moving the line to at least the first round, so the line moves to include first-round losers in the lottery.

Most years, the line should not need to be moved at all. In years that it is moved, it is unlikely to be moved beyond the first round of the playoffs. Intentionally losing in the playoffs is not something we have seen before. It is unlikely in any case, and with this safeguard it can be prevented entirely. 

When the line is moved to include playoff teams in the lottery, lottery index increments depend on how far the line moves. If the line moves to include first-round losers, all teams in the lottery receive the standard $\alpha$ increment, as eight teams remain excluded and the distinction between lottery and non-lottery teams still matters. However, if the line moves beyond the second round to include second-round losers or teams that advanced further, no teams receive increments that year. At this threshold, only four or fewer teams are excluded from the lottery, making it nearly universal. Lottery odds are then determined purely by accumulated multi-year performance. In this situation, the previous year's champion, whose lottery index was reduced to zero, cannot win the lottery. Recent playoff teams have diminished lottery indices from their success, giving them low odds. Persistently weak teams retain their accumulated tickets, maintaining their advantage.

Teams included in the lottery by moving the line are eligible to win top draft picks. If they win a top draft pick, their lottery index is diminished according to which pick they received, just as it would be for any lottery winner. Playoff success up to the point where the line is drawn does not diminish their lottery index that year.

\subsubsection{Implementation of the Bayesian Truth Serum}

In order to implement RBTS in this context, it is necessary to specify an information score and a prediction score. The information score is the reward that agents receive for their own response. The prediction score is the reward that agents receive for their prediction of the poll results.

For the prediction score, we follow Radanovic and Faltings \cite{radanovic2013nonbinary} and use a quadratic scoring rule that compares an agent's prediction against a peer's report. Let $\mathbf{y}_a$ be the prediction vector for agent $a$, where $\mathbf{y}_a(i)$ is the probability agent $a$ predicts for option $i$. Let $x_j$ denote the option chosen by peer $j$. The score for agent $a$ with respect to peer $j$ is:

\begin{equation}
R(\mathbf{y}_a, x_j) = \frac{1}{2} + \mathbf{y}_a(x_j) - \frac{1}{2} \sum_{i=1}^{6} \mathbf{y}_a(i)^2
\end{equation}

To avoid the all-or-nothing extreme of pairing with just one random peer, the final prediction score is the average over all peers:

\begin{equation}
\text{Prediction Score}_a = \frac{1}{n-1} \sum_{j \neq a} R(\mathbf{y}_a, x_j)
\end{equation}

The prediction score is maximized when agents truthfully report their beliefs about the distribution of responses.

For the information score, Radanovic and Faltings \cite{radanovic2013nonbinary} provide a 1-peer incentive-compatible information score. A 1-peer score results in rewards that are very unpredictable (all or nothing), since the reward is based on which peer an agent is randomly paired with. On the other hand, if an agent is paired against all other agents, then any two agents will know that they will be compared with each other and rewarded for matching, which makes collusion more likely. To avoid both extremes, we use a compromise where each agent is compared with one fourth of the agents, randomly selected while excluding conflicts of interest (agents who cover the same team or work for the same organization). The choice of one-fourth is large enough to smooth the information score across multiple comparisons, yet small enough that agents cannot reliably anticipate or coordinate with their comparison group. The information score is:

\begin{equation}
\text{Information Score}_a = \frac{1}{n/4} \sum_{j \in \text{peers}} \frac{1}{\mathbf{y}_j(x_a)} \cdot \mathbf{1}_{x_j = x_a}
\end{equation}

where $n$ is the total number of agents, the sum is taken over the $n/4$ randomly selected peers, and $\mathbf{1}_{x_j = x_a}$ is an indicator variable equal to 1 when the peer's response matches the agent's response.

The prediction score rewards agents for accurately forecasting what others will report, incentivizing honest predictions. The information score creates a more subtle incentive: the numerator increases when the agent's response matches their peers, while the denominator increases when peers predict the agent's chosen response is rare. An agent maximizes this ratio by reporting their true belief, because doing so best aligns their response with what they genuinely expect others to report. Together, these scores make honesty the optimal strategy. For full technical details, see Radanovic and Faltings \cite{radanovic2013nonbinary}.

To ensure well-defined scores, predictions must be non-zero. We set a minimum of 0.1\% probability for each option, so $\mathbf{y}_a (i) \geq 0.001$ for all $i$, with $\sum_{i=1}^{6} \mathbf{y}_a (i) = 1$.

Agents are compensated from a fixed league-funded budget. There is little to no external reason for the analysts to be dishonest, so the budget only needs to be large enough to incentivize participants to invest the effort required to form and defend their true assessment. Compensation is essentially payment for their time, though the distribution will be uneven based on the mechanism's scoring. The payment is based on an even split between prediction and information scores:

\begin{equation}
\begin{aligned}
\text{Agent Payment} 
&= \frac{\text{Agent's Prediction Score}}{\sum_{\text{all agents}} \text{Prediction Score}} \times \frac{\text{Budget}}{2} \\
&\quad + \frac{\text{Agent's Information Score}}{\sum_{\text{all agents}} \text{Information Score}} \times \frac{\text{Budget}}{2}
\end{aligned}
\end{equation}

This structure ensures that both components contribute equally to the total compensation regardless of their raw magnitudes.

\subsection{Parameters and Variations}

There are three facets to COLA's core structure: no benefit for additional losses after playoff elimination, carry-over, and diminishment. There are numerous variations and parameters that can be tuned within that framework. These choices can suit different league priorities without affecting incentive compatibility:

\textbf{Increment Size.} Classic COLA uses an annual increment of $\alpha = 1000$, but a different value could be used. To maintain incentive compatibility, it is important that a team not gain any benefit from additional losses, but it is not necessary for $\alpha$ to be constant. For example, the increment could be larger for teams that have missed the playoffs for multiple years in a row. It could also be based on other factors, even giving teams additional lottery tickets based on how many games they win. As long as losses do not earn more lottery tickets, incentive compatibility is maintained.

\textbf{Diminishment Rates.} Classic COLA's design (100\%, 75\%, 50\%, 25\% for draft picks; similar for playoff rounds) prioritizes simplicity. More aggressive rates emphasize recency; gentler rates give historical performance more weight. More aggressive rates on playoff success also increase the likelihood that the line must be moved to maintain incentive compatibility. More complex diminishment rules could be employed. Beyond percentages, diminishment could use fixed amounts, caps/floors, or escalation rules (e.g., doubled rates for teams making three consecutive playoff appearances).

\textbf{Lottery Scope.} Following current NBA practice, the lottery for Classic COLA is for the top four picks, but this extends naturally to more or fewer positions if needed. The mechanism works identically with any number of lottery positions, though more positions increase the potential for the worst team to receive a poor pick despite having the best odds. 

\textbf{Non-Lottery Pick Assignment.} Picks 5-14 could be assigned by current-season standings (matching the existing NBA system) or by lottery index (consistent with COLA's multi-year evaluation). Standings-based assignment provides familiarity, while lottery-index assignment rewards historically weak teams throughout the draft order, not just through the lottery. The latter also mitigates the incentive to tank for the purpose of retaining a traded draft pick that has protections. We recommend banning those sorts of trade protections if current-season standings are used to assign the non-lottery picks.

\textbf{No Lottery At All.} If the draft pick order is determined by ordering all teams by highest lottery index (as suggested in the previous variant), then Playoff Track Record and prior drafts would determine the draft order instead of regular season standings. The increment rule, carry-over, and diminishment for playoff success and top draft picks accomplish their goals \textit{with} or \textit{without} a lottery. (Diminishment for top picks should remain, even if not determined by lottery. Otherwise, the same team would pick first each year until making the playoffs.) However, the lottery also generates fan interest, so preventing tanking is not its only purpose.

\textbf{Opt-Out Penalty.} Classic COLA's $2\alpha$ cost for skipping weak draft classes balances individual flexibility against the desire for league-wide participation. Lower the value to increase team autonomy. Adjust higher or eliminate the option if more participation is desired.

\textbf{Moving the Line.} Our six-point scale on the analyst survey could be expanded to more options for more granularity at the expense of complexity for survey participants. For example, an alternative design could expand the survey to twelve options by allowing each playoff round to be included in the lottery either \textit{with} or \textit{without} diminishment for playoff success at that level.

Pros and cons for having the league office or a Blue Ribbon Commission decide how to set the line between lottery and non-lottery teams were discussed in Section \ref{sec:StrongDraft}. Alternatively, the line could be set as immovable. A static line is simpler, but it risks introducing a tanking hazard (if placed too low) or not providing enough help to weak teams (if placed too high). 

\textbf{Increments in Strong Draft Years.} When the line moves to include playoff teams, Classic COLA uses a binary rule: if eight or more teams remain excluded, all lottery participants receive $\alpha$; if four or fewer remain excluded, nobody receives $\alpha$. A simpler alternative would always grant $\alpha$ to all lottery participants regardless of where the line falls, prioritizing ease of explanation over maximizing Preferencing for Quality in strong draft years.

\textbf{Traded Picks Winning the Lottery.} Under Classic COLA, traded draft picks are excluded from lottery eligibility. Alternative designs could allow traded picks to enter the lottery, with an appropriate lottery index diminishment rule. For example, diminishment could be applied to both the pick holder and the original team if the pick wins. This would introduce a third class of traded draft picks: protected 1–4, unprotected but lottery-ineligible, and unprotected lottery-eligible picks. While this variant improves trade-market liquidity and gives teams more flexibility, it is substantially more complex and creates the opportunity for systematic mismanagement of long-run lottery risk by short-term decision-makers.

\section{Simulation} \label{sec:Simulation}

We present simulation results to further demonstrate that, under COLA, weaker teams eventually have the opportunity to improve and compete with stronger teams. The results indicate that no team appears permanently locked into a top or bottom position in the league by the draft mechanism. 

The simulations ignore differences in front office skill, media market size, and other factors that will result in real-world variation in team success, including the possibility of teams spending the long-term as either a winning or losing team. These results demonstrate that any such variation does not come from the draft lottery mechanism.

\subsection{Simulation Description}

This simulation models the long-run behavior of a league operating under COLA, with the goal of validating that the mechanism behaves as intended. In particular, it tracks how team performance, draft outcomes, and accumulated lottery indices evolve over time, and checks that COLA does not generate persistent dominance or runaway advantage. The simulation is not intended to discover new theoretical results, but to provide a basic robustness check that the proposed mechanism produces sensible outcomes under stylized league dynamics. The simulation assumes the standard case where only non-playoff teams participate in the lottery; strong draft-year line adjustments are not simulated.

\subsubsection{League Structure}

The league consists of \(N = 30\) teams. Each team \(i\) is characterized by two state variables: Team strength ($S_i^t$) and Lottery index ($L_i^t$).

Initial team strength is drawn uniformly from 5 to 100. Weaker teams are assigned higher initial lottery indices, reflecting the empirical tendency for weaker teams to enter the lottery with a higher lottery index.

Each team plays a total of 82 games in the regular season. Matchups are generated such that every pair of teams plays at least two games against each other, and the remaining games are assigned randomly.

\paragraph{Game Outcomes}

For a game between teams \(i\) and \(j\), the probability that team \(i\) wins is based on the Bradley–Terry model \cite{bradley1952rank}:

\[
P(i \text{ wins}) =
\frac{S_i}{S_i + S_j} \quad 
\] 

Teams are ranked by total regular-season wins. The top 16 teams qualify for the playoffs. Conference and division distinctions are ignored, but the playoffs otherwise follow standard NBA best-of-seven single-elimination rules.

\subsubsection{Draft Lottery}

The draft lottery is implemented according to COLA:

\begin{itemize}
    \item Each team receives exactly one draft pick per season.
    \item Only non-playoff teams participate in the lottery, with selection probability proportional to the number of accumulated lottery tickets.
    \item Lottery indices are updated based on playoff results and draft results.    
\end{itemize}

Following the lottery, teams are assigned draft positions, where each position \(p_i\) is associated with a coefficient \(c(p_i)\) representing the expected long-term value of a player selected at that pick. These coefficients are derived from empirical analysis of NBA draft outcomes \cite{statheadBasketball} and are sorted in descending order, so that higher picks correspond to greater expected value while later picks provide diminishing returns.

\subsubsection{Team Strength Dynamics}

Team strength evolves through three sequential mechanisms applied at the end of every season: decay, draft-based improvement, and redistribution.

\paragraph{Strength Decay}

At the conclusion of each season, team strength declines due to aging, injuries, and roster turnover, following a multiplicative decay similar to the relaxation of agent status in the Bonabeau hierarchy model~\cite{bonabeau1995phase}:

\[
S_i \leftarrow S_i (1 - d_i)
\]

where:

\[
d_i \sim \mathcal{U}(0.05, 0.15)
\]

\paragraph{Draft-Based Strength Increase}

After the decay process, team strength is increased based on the team’s assigned draft position following a partial adjustment approach inspired by the Rescorla--Wagner model \cite{rescorla1972}:

\[
\Delta S_i = c(p_i) \cdot (S_{\max} - S_i) \cdot \beta
\]

where \(c(p_i)\) is the draft coefficient associated with pick \(p_i\), \(S_i\) is the team’s current strength, \(S_{\max}\) is the maximum attainable strength, and \(\beta\) is a global scaling parameter.

The scaling factor is set to \(\beta = 7.5\), which balances with the decay to produce the most stable and realistic long-run outcomes in the simulation. For large values of \(\beta\), multiple teams hit the maximum team rating, becoming indistinguishable. For small values, the draft-based increase is outweighed by the decay, resulting in one team remaining consistently strong while the others gradually weaken. This formulation ensures that draft benefits are larger for weaker teams, while diminishing as team strength approaches the maximum.

\paragraph{Strength Spreading}

Strengths are redistributed after decay and draft-based improvement. The new lower and upper bounds, \(S_{\min}^{\text{new}}\) and \(S_{\max}^{\text{new}}\), are taken as the minimum and maximum of \(N = 30\) independently drawn random numbers within the strength range:

\[
S_{\min}^{\text{new}},\; S_{\max}^{\text{new}} \sim \mathcal{U}(S_{\min}, S_{\max})
\]

Team strengths are then linearly scaled to fit within these bounds, preserving relative rankings while maintaining realistic dispersion.

\subsection{Simulation Results}

\subsubsection{Balance}

\begin{figure}[h]
    \centering
    \includegraphics[width=0.7\textwidth]{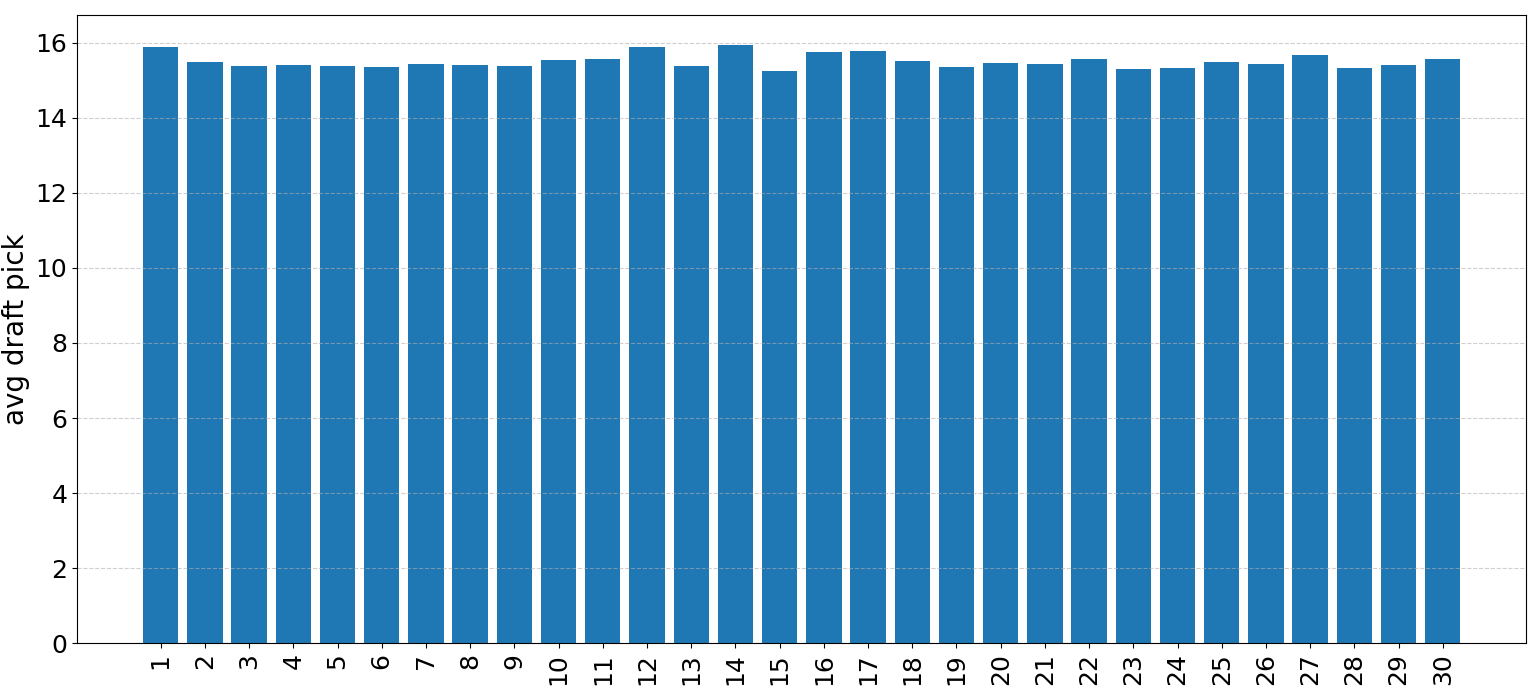}
    \caption{Average draft pick after 1000 seasons for 30 teams}
    \label{fig:avg_draft_pick}
\end{figure}

Figure \ref{fig:avg_draft_pick} shows the average draft pick for each of the 30 teams across a simulation of 1000 seasons. On average, over a long time period all teams had an average draft pick of around 15.

\begin{figure}[h]
    \centering
    \includegraphics[width=0.7\textwidth]{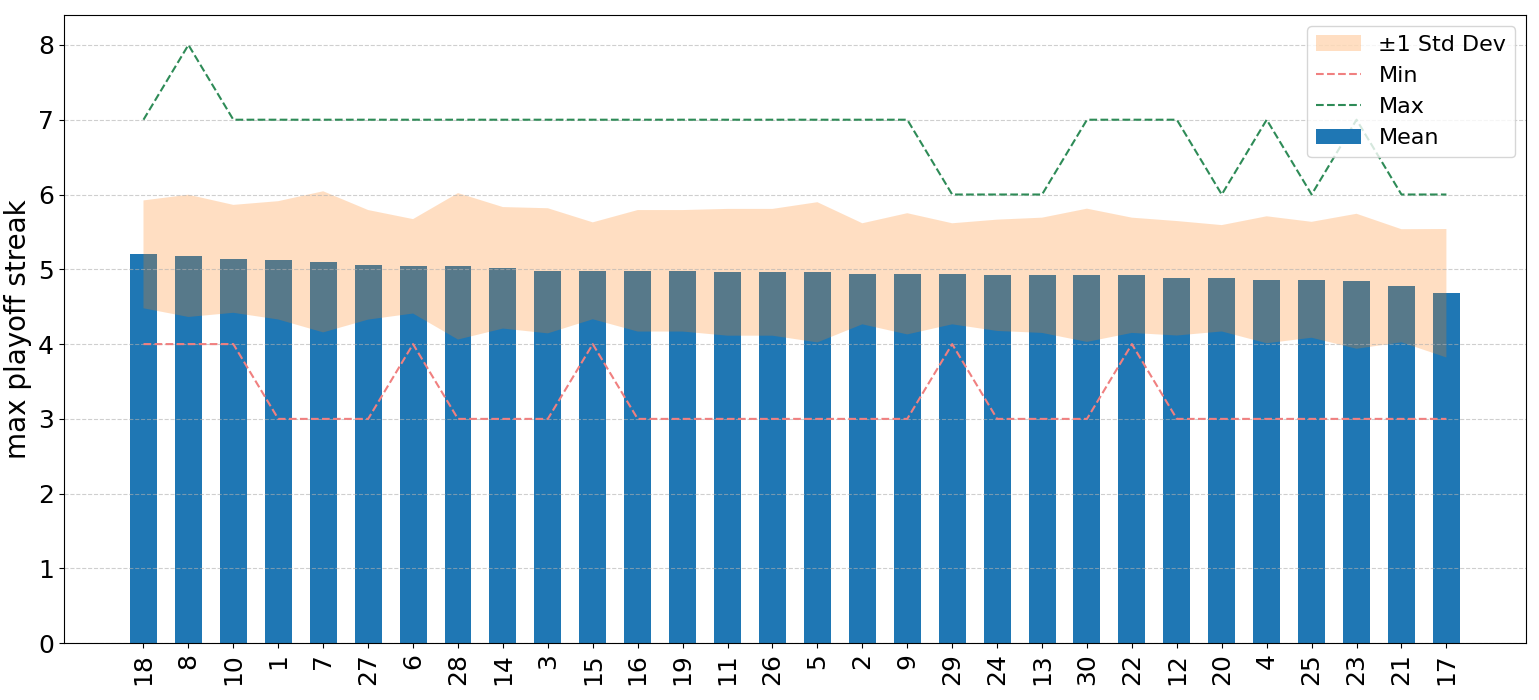}
    \caption{Longest playoff streak, 50 simulations of 100 seasons each}
    \label{fig:longest_playoff_streak}
\end{figure}

\begin{figure}[h]
    \centering
    \includegraphics[width=0.7\textwidth]{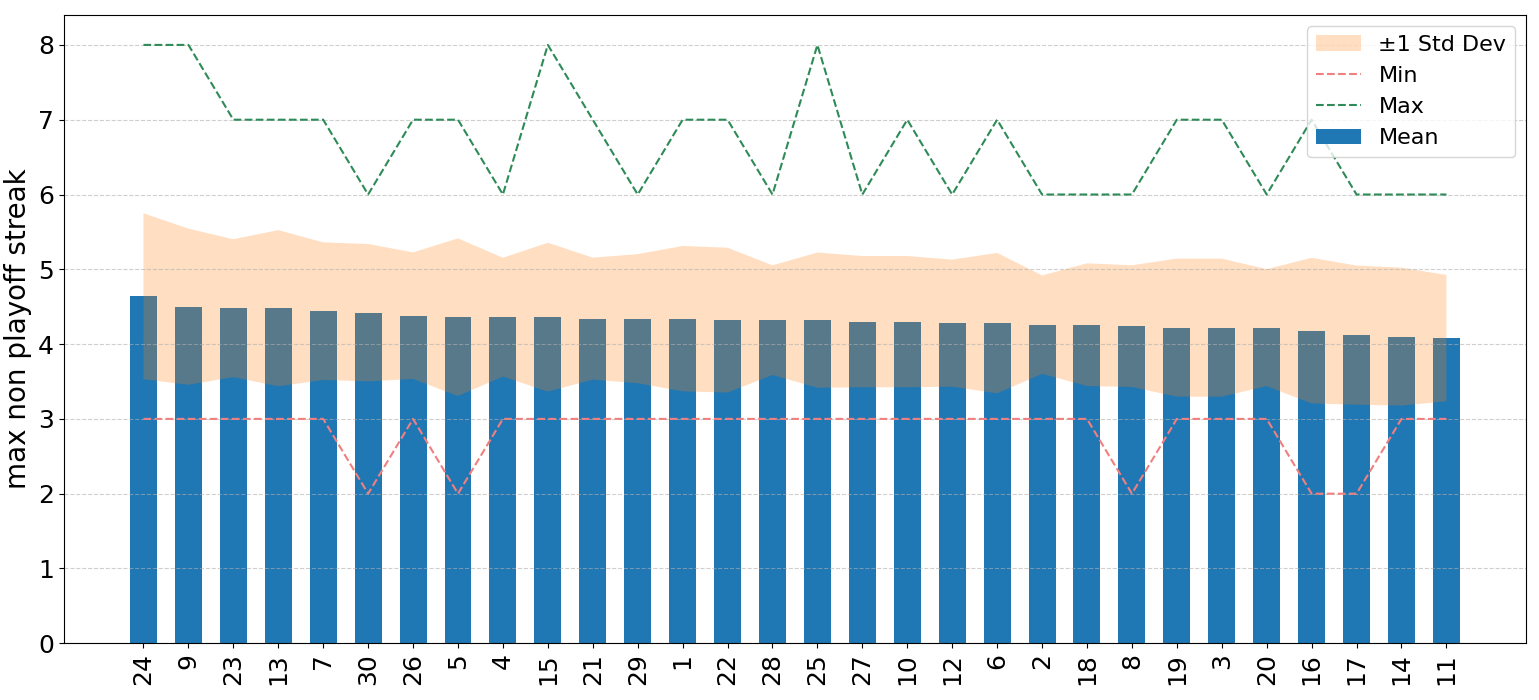}
    \caption{Longest streak missing playoffs, 50 simulations of 100 seasons each}
    \label{fig:longest__non_playoff_streak}
\end{figure}

Figures \ref{fig:longest_playoff_streak} and \ref{fig:longest__non_playoff_streak} are based on the Monte Carlo experiment consisting of 50 simulations of 100 seasons each. Figure \ref{fig:longest_playoff_streak} shows, for each team, the longest streak of consecutive seasons making the playoffs. Figure \ref{fig:longest__non_playoff_streak} shows the longest streak of consecutive seasons missing the playoffs. Even over a long time period and 50 iterations, no team missed the playoffs more than ten years in a row. These results provide evidence that the mechanism does not lock teams permanently into success or failure.

\begin{figure}[h]
    \centering
    \includegraphics[width=0.7\textwidth]{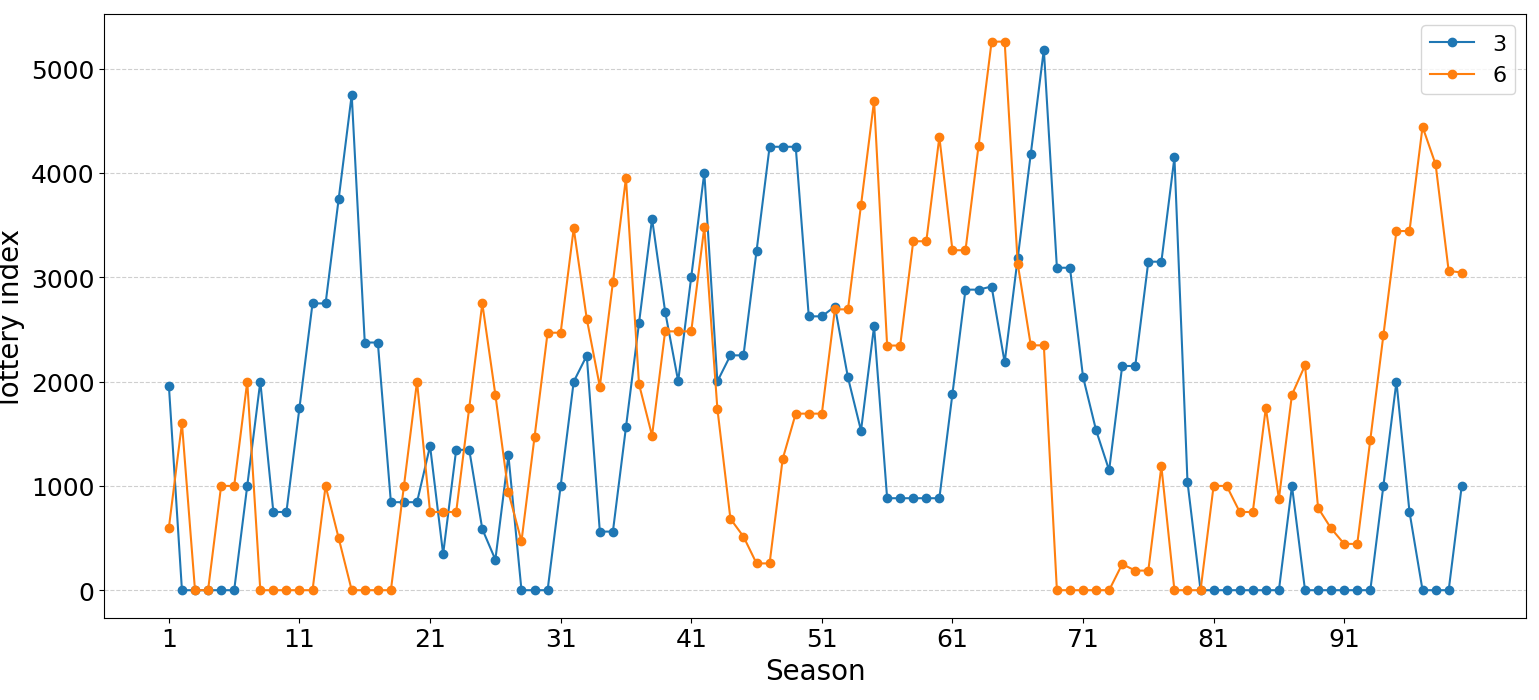}
    \caption{Tracking the lottery index for 2 teams over 100 seasons}
    \label{fig:lottery_index}
\end{figure}

Figure \ref{fig:lottery_index} shows the evolution of lottery indices over 100 simulated seasons for two randomly selected teams. The indices fluctuate in response to seasonal performance, playoff outcomes, and draft positions. These dynamics illustrate that COLA creates the intended effect: teams alternate between periods of high lottery index and periods of low lottery index, and teams do not consistently dominate or stay at the bottom.

\subsubsection{Losing to Exclude Others from the Lottery}

Section \ref{sec:Assumptions} noted that a team might have an incentive to lose if doing so swaps a high-index opponent into the playoffs in exchange for a low-index opponent entering the lottery. We analyzed this scenario across a 1000-season simulation.

On average, the pool of lottery tickets for a draft contained 43,410 tickets. We examined the teams on either side of the playoff line: the two lowest-index playoff teams and the two highest-index non-playoff teams. The average difference between the higher-index playoff team and the lower-index non-playoff team was 1805 tickets. This represents a conservative (i.e. high) estimate of the maximum potential benefit if a team successfully manipulates the standings to swap two teams.

The non-playoff team with the highest lottery index stands to gain the most. The average high was 6015 tickets, so successfully executing this strategy would improve their win probability from 6015/43410 = 13.9\% to 6015/(43410 - 1805) = 14.5\%, a gain of 0.6\%. Over 1000 seasons, the greatest potential improvement was 3.0\%. In over 90\% of the seasons, the greatest potential improvement any team might have had was less than 1.5\%.

Thus, the potential benefit a team could achieve by manipulating playoff qualification is too low to be a factor in practice.

\section{Conclusion} \label{sec:Conclusion}

We have presented the Carry-Over Lottery Allocation (COLA) Draft Mechanism, a draft mechanism that is incentive-compatible (anti-tanking), advantages the worst teams, and is practical for real-world implementation. COLA circumvents the fundamental impossibility proven by Munro and Banchio \cite{munro_banchio} through a paradigm shift: evaluating team quality based on multiple years of playoff performance instead of single-season regular season record. 

COLA itself is our primary contribution. It is a mechanism where each team maintains a lottery index that tracks accumulated lottery tickets. All non-playoff teams receive equal annual increments to their lottery index, tickets can carry over to future seasons, and playoff success or lottery wins diminish the index according to achievement level. This is, to our knowledge, the first draft mechanism to simultaneously meet these five desirable criteria:

\begin{enumerate}
    \item \textbf{Practicality}: COLA uses a familiar draft lottery that is easy for fans to follow. The mechanism does not require unrealistic behavioral assumptions or precise measurements of unknowable team preferences.
    \item \textbf{Anti-Tanking}: All non-playoff teams receive equal lottery index increments, eliminating any advantage for losing additional games.
    \item \textbf{Preferencing for Quality}: Success reduces a team's lottery index, favoring persistently weak teams.
    \item \textbf{Preferencing for Luck}: Lottery wins reduce a team's lottery index, favoring teams with worse historical lottery outcomes.
    \item \textbf{Playoff Track Record}: Multi-year playoff results are used to measure team quality instead of single-season standings.
\end{enumerate}

These criteria hold under reasonable assumptions about team incentives, such as preferring playoff qualification over lottery participation. When the assumptions fail, mitigation mechanisms preserve incentive compatibility in practice.

The implementation challenges detailed in Section \ref{sec:Implementation} demonstrate COLA's feasibility. Traded draft picks are handled in a way that preserves the mechanism's core properties. To transition to the system, historical data from 1999 onward establishes fair starting values for lottery indices. When there is a weak draft class, a team is allowed to opt out for a cost, while strong draft classes are identified through an annual media survey that may expand lottery access. Each solution preserves incentive compatibility while accommodating the complexities of actual league operations. The framework is robust to myriad parameter choices and variants, which can be used to tune the Preferencing for Luck and Preferencing for Quality to the desired levels. Leagues may debate the optimal ways to adjust increment sizes, diminishment rates, lottery scope, and more while preserving incentive compatibility.

COLA demonstrates that the tension between rewarding weak teams and eliminating tanking incentives is resolvable. By evaluating team quality through multi-year playoff performance and treating all non-playoff teams equally each season, the mechanism achieves both goals. The COLA Draft Mechanism provides a solution to a problem that has plagued professional basketball for decades.

\appendix

\section{Appendix: Proof of the Existence of a Draft Mechanism that is Incentive-Compatible and Favors the Worst Team}

Munro and Banchio's impossibility theorem leads us to ask whether, with a different definition of \textit{worst team}, there exists an incentive-compatible draft mechanism that favors the worst team. Here, we prove that, under a reasonable definition of worst team and set of assumptions about team preferences, such a mechanism does exist.

\subsection{League Structure}

We consider a league with 30 teams and a standard playoff structure of 16 playoff teams and 4 playoff rounds. Seeding, conference differences, and home-court advantage are ignored. After the playoffs, there is a draft of 22 incoming players and a lottery to determine which teams will have the right to make those selections. (There may also be other draft picks beyond those 22, but they are assumed to have negligible value and are not factored into the model.)

\subsection{Team Preferences and Definition of Worst Team}

We define the \textit{worst team} as the team with the longest active streak of seasons without winning a playoff series.

Each team maximizes an objective function with the following priorities:

\begin{enumerate}
    \item Primary goal: playoff series wins $s$
    \item Secondary goal: draft picks $d_1 > d_2 > \dots > d_{22}$
    \item Tertiary goal: regular season game wins $g$
\end{enumerate}

In other words, a team's preferences satisfy:
\[
\text{Primary} \;>\; \text{Secondary} \;>\; \text{Tertiary} \;\gg\; \text{other considerations}.
\]

Let:

\begin{itemize}
    \item $x_s \in \{0,1,2,3,4\}$ denote the number of playoff series won,
    \item $x_k \in \{0,1\}$ denote an indicator for receiving draft pick $k$,
    \item $x_g \in \{0,1,\dots,82\}$ denote the number of regular-season games won.
\end{itemize}

Utility is given by

\begin{equation}
U = x_s s + \sum_{k=1}^{22} x_k d_k + x_g g,
\end{equation}

where $ s > d_1 > d_2 > \dots > d_{22} > g > 0$.

For each game, a team chooses to either try to win or to lose. Games are independent, which means that if a team tries to win, tries to lose, does win, or does lose a game, it has no impact on the team's ability to win a future game.

\subsection{Diet COLA Mechanism}

We define Diet COLA as a variant of COLA where each team $t$ has a lottery index $L_t$ that carries over from year to year.

\begin{itemize}
    \item If a team does not win a playoff series, then the team's lottery index is unchanged for this year's lottery. It will be incremented by one for the next year:
    \begin{equation}
    L_t' = L_t + 1.
    \end{equation}
    
    \item If a team wins a playoff series, then the team's lottery index is reduced to zero:
    \begin{equation}
    L_t' = 0.
    \end{equation}
\end{itemize}

All teams that fail to win a playoff series participate in the lottery, and 22 draft picks are distributed via lottery.

Each team $t$ that participates in the lottery has a number of tickets equal to $L_t$.

Draft pick trades are banned.

\subsection{Pool Manipulation Assumption}

The probability that team $t$ wins the first draft pick is:
\begin{equation}
\text{P(\#1 pick)} = \frac{L_t}{\sum_{j \in \text{lottery}} L_j}.
\end{equation}

A team cannot affect $L_t$ other than the standard increment that occurs if they fail to win a playoff series. Since $s > d_t$, it is never better to lose in an attempt to increase $L_t$. 

However, a team could attempt to change the denominator by manipulating the pool of lottery opponents. Suppose a team would lose to a high-index opponent, pushing that opponent into the playoffs, and then that high-index opponent would beat a low-index team that falls into the lottery. That would change the denominator, increasing the team's probability of getting the number one pick.

There are several reasons why we do not expect teams to employ this technique. 

\begin{itemize}
    \item Until it is late in the season, it is unknown whether or not a high-index team will end up near the playoff border where this sort of technique would even make a difference.
    \item If a high-index team is successfully pushed into the playoffs, there is no guarantee that they win a playoff series, so they may end up in the lottery anyway.
    \item If a high-index team does win a playoff series, the benefit is limited by the index of the team that replaces them.
    \item If the technique is successful, the benefit of having a higher probability of getting a top pick is spread out among all the teams in the lottery, weighted toward the teams with the most tickets. Because of this, even a large swing in the size of the ticket pool means only a small gain for any individual team.
    \item The nature of the draft mechanism will tend to minimize the maximum drought length, which also limits the potential benefit of this technique.
    \item Normal tanking requires a team to be weak in general, but this technique requires losing a specific game, which is a different sort of intentional losing. This sort of targeted losing is akin to throwing a game for the purposes of gambling, and penalties for the latter could include a permanent ban from the league for anyone involved. 
\end{itemize}

Simulations in Section \ref{sec:Simulation} indicate that in 90\% of seasons, this technique would not offer the opportunity for a significant probability increase. Under Diet COLA, where a team could lose in the playoffs but still be in the lottery, the technique would be even less effective. 

Because of the limited and uncertain effectiveness of this technique and the potentially severe penalties if discovered, we make the reasonable assumption that teams value winning the game for its own sake more than the expected benefit of manipulating the pool of opponents in the lottery:

\begin{equation}
g > \mathbb{E}[\text{gain from opponent manipulation}].
\end{equation}

\subsection{Probability of Playoffs Wins and Lottery Participation}

Note that if a team participates in the lottery, they will receive a draft pick that can be treated as a fixed value, $d_t$. Though the specific draft position is unknown until the lottery, teams are unable to affect which draft pick they will receive this year because they cannot improve $L_t$. Thus each team $t$ will either get draft pick with value $d_t$ or win a playoff series.

Let:
\begin{itemize}
    \item $P(s)$ = probability of eventually winning a playoff series
    \item $P(d_t)$ = probability of remaining in the lottery and receiving draft value $d_t$
\end{itemize}

Because these are complementary outcomes,

\begin{equation}
P(s) + P(d_t) = 1.
\end{equation}

Thus,

\begin{equation}
\Delta P(d_t) = - \Delta P(s).
\end{equation}

\subsection{Incentive Compatibility Proof}

We compare the utility impact of winning versus losing a game.

\subsubsection*{Case 1: Before Elimination}

If the team wins, the marginal utility is
\begin{align}
U_W 
&= s \,\Delta_W P(s) + d_t \,\Delta_W P(d_t) + g \\
&= s \,\Delta_W P(s) + d_t \left(-\,\Delta_W P(s)\right) + g \\
&= \Delta_W P(s)\,(s - d_t) + g,
\end{align}

If the team loses:

\begin{align}
U_L 
&= s \,\Delta_L P(s) + d_t \,\Delta_L P(d_t) + 0 \\
&= s \,\Delta_L P(s) + d_t \left(-\,\Delta_L P(s)\right) \\
&= \Delta_L P(s)\,(s - d_t).
\end{align}

Because
\[
s > d_t, \quad g > 0, \quad \Delta_W P(s) > 0, \quad \text{and} \quad \Delta_L P(s) < 0,
\]
it follows that
\[
U_W > 0 \quad \text{and} \quad U_L < 0.
\]

Therefore, it is strictly better to try to win.

\subsubsection*{Case 2: After Elimination}

After elimination:

\begin{equation}
P(s) = 0, \quad P(d_t) = 1.
\end{equation}

Winning yields:

\begin{equation}
U_W = g.
\end{equation}

Losing yields:

\begin{equation}
U_L = 0.
\end{equation}

Since $g > 0$, winning is strictly preferred.

\subsubsection*{Case 3: After Clinching Playoffs}

In this case:

\begin{equation}
\Delta P(s) = 0, \quad \Delta P(d_t) = 0.
\end{equation}

Winning yields $g$, losing yields $0$, so again winning is strictly preferred.

\subsubsection*{Case 4: During the Playoffs}

In the first playoff round:

\begin{align}
U_W &= \Delta_W P(s)(s - d_t), \\
U_L &= \Delta_L P(s)(s - d_t).
\end{align}

Since $\Delta_W P(s) > 0$ and $\Delta_L P(s) < 0$, winning strictly dominates losing. (In this model, $s$ represents the inherent value of winning a playoff series, but no value has been assigned to individual playoff games. If a value were assigned to the individual games, it would not change the analysis.)

\subsection{Main Result}

Under Diet COLA, no team ever prefers to lose a game. The team with the longest active streak without a playoff series win has the highest lottery index. Thus, the highest lottery index implies the highest probability of receiving the first draft pick.

The model includes these assumptions:
\begin{enumerate}
    \item Teams prefer to win a playoff series over getting any draft pick.
    \item Teams prefer higher draft picks to lower draft picks.
    \item Teams prefer any draft pick to winning any single game.
    \item Teams prefer winning a game to other factors (which are considered in this model).
    \item Teams prefer winning a game to the uncertain and limited benefit of attempting to manipulate the pool of opponents in the lottery. 
\end{enumerate}

Our main result is that under these assumptions, a simplified version of the Carry-Over Lottery Allocation draft mechanism (Diet COLA) is provably incentive-compatible, and under Diet COLA the team with the longest active streak of seasons without a playoff series win has the highest probability of receiving the first draft pick.

In other words, under reasonable assumptions, the Diet COLA draft mechanism is an anti-tanking draft mechanism that favors the worst team.

\medskip

\printbibliography

\end{document}